\documentclass{PoS}
\pdfoutput=1
\usepackage[utf8]{inputenc}
\usepackage{cite}
\usepackage[hyphens]{url} 
\usepackage[safe]{textcomp}
\usepackage{bbm} 
\usepackage{amstext}  
\usepackage{amsfonts}
\usepackage{amsthm}
\usepackage{graphicx}
\usepackage{amsmath}
\usepackage{amsfonts}
\usepackage{mathtools} 
\usepackage{array}                
\usepackage{xcolor}  
\usepackage[absolute]{textpos}
\usepackage{multirow} 
\usepackage{slashed}
\usepackage{relsize}

\newcommand{\OpenLoops}{\text{\sc OpenLoops}}
\newcommand{\OneLoop}{\text{\sc OneLoop}}
\newcommand{\Collier}{\text{\sc Collier}}
\newcommand{\Cuttools}{\text{\sc Cuttools}}
\newcommand{\p}{\partial}
\newcommand{\f}[2]{\frac{#1}{#2}}

\newcommand{\ssst}[1]{\scriptscriptstyle{\text{#1}}}
\newcommand{\nosss}[1]{#1}

\newcommand{\bit}{\begin{itemize}}
\newcommand{\eit}{\end{itemize}}
\newcommand{\bce}{\begin{center}}
\newcommand{\ece}{\end{center}}
\newcommand{\bea}{\begin{eqnarray}}
\newcommand{\eea}{\end{eqnarray}}
\newcommand{\be}{\begin{equation}}
\newcommand{\ee}{\end{equation}}
\newcommand{\ba}{\begin{align}}
\newcommand{\ea}{\end{align}}
\newcommand{\beas}{\begin{eqnarray*}}
\newcommand{\eeas}{\end{eqnarray*}}
\newcommand{\bes}{\begin{equation*}}
\newcommand{\ees}{\end{equation*}}
\newcommand{\bas}{\begin{align*}}
\newcommand{\eas}{\end{align*}}
 
 \newcommand{\Rig}{\Rightarrow}
 
\newcommand{\eps}{{\varepsilon}}

\newcommand{\lb}{\left(}
\newcommand{\rb}{\right)}

\newcommand{\idop}{1\!\!\!1}

\newcommand{\Dbar}[1]{\bar{D}_{\nosss{#1}}}

\newcommand{\momp}[1]{p_{#1}}
\newcommand{\mass}[1]{m_{\nosss{#1}}}

\newcommand{\momq}{\bar{q}}
\newcommand{\tilq}{\tilde{q}}

\newcommand{\calA}{\mathcal{A}}

\newcommand{\calM}{\mathcal{M}}
\newcommand{\calN}{\mathcal{N}}

\newcommand{\calW}{\mathcal{W}}

\newcommand{\seg}{S}

\newcommand{\col}{\mathrm{col}}
\newcommand{\hel}{\mathrm{hel}}
\newcommand{\re}{\mathrm{Re}}
\newcommand{\Tr}{\mathrm{Tr}}

\newcommand{\rd}{\mathrm d}

\definecolor{bluemar}{rgb}{0,0,.5}
\definecolor{redmar}{rgb}{.8,0,0}
\definecolor{greenmar}{rgb}{0,.5,0}

\newcommand{\deltathr}{\delta_{\mathrm{thr}}}

\title{On-the-fly reduction of open loops}

\ShortTitle{On-the-fly reduction of open loops}

\author{Federico Buccioni\\
        University of Zurich, Zurich, SWITZERLAND\\
        E-mail: \email{buccioni@physik.uzh.ch}}
        
\author{Jean-Nicolas Lang\\
        University of Zurich, Zurich, SWITZERLAND\\
        E-mail: \email{jlang@physik.uzh.ch}}              
        
\author{Stefano Pozzorini\\
        University of Zurich, Zurich, SWITZERLAND\\
        E-mail: \email{pozzorin@physik.uzh.ch}}
        
 \author{Hantian Zhang\\
        University of Zurich, Zurich, SWITZERLAND\\
        E-mail: \email{hantian.zhang@physik.uzh.ch}}         
        
\author{\speaker{Max Zoller}\\
       University of Zurich, Zurich, SWITZERLAND\\
       E-mail: \email{zoller@physik.uzh.ch}} 

\abstract{We describe new developments in the {\sc OpenLoops} framework based on the recently introduced on-the-fly 
method \cite{Buccioni:2017yxi}.
The on-the-fly approach exploits the factorisation of one-loop diagrams into segments in order to perform various operations, 
such as helicity summation, diagram merging
and the reduction of Feynman integrands in between the recursion steps for the amplitude construction. 
This method significantly reduces the complexity of scattering amplitude calculations for multi-particle processes, 
leading to a major increase in CPU efficiency and numerical stability.
The unification of the reduction to scalar integrals with the amplitude construction in a single algorithm, 
allows to identify problematic kinematical configurations 
and cure numerical instabilities in single recursion steps. 
A simple permutation trick in combination with a one-parameter expansion for a single topology, 
which is now implemented to any order,
eliminate rank-two Gram determinant instabilities altogether. Due to this any-order expansion, 
the numerical accuracy of the algorithm can be determined with a rescaling test.
The on-the-fly algorithm is fully implemented for double and quadruple precision, 
which allows for true quadruple precision benchmarks with up to 32 correct digits as well as
a powerful rescue system for unstable points. 
We present first speed and stability results for these new features.
The on-the-fly algorithm is part of the forthcoming release of {\sc OpenLoops 2}.}

\FullConference{Loops and Legs in Quantum Field Theory (LL2018)\\
		29 April 2018 - 04 May 2018\\
		St. Goar, Germany}

\begin{document}

\section{Automated amplitude generation in OpenLoops{}}
\begin{figure}[t!]\begin{center} 
$\parbox{8cm}{\includegraphics[height=65mm]{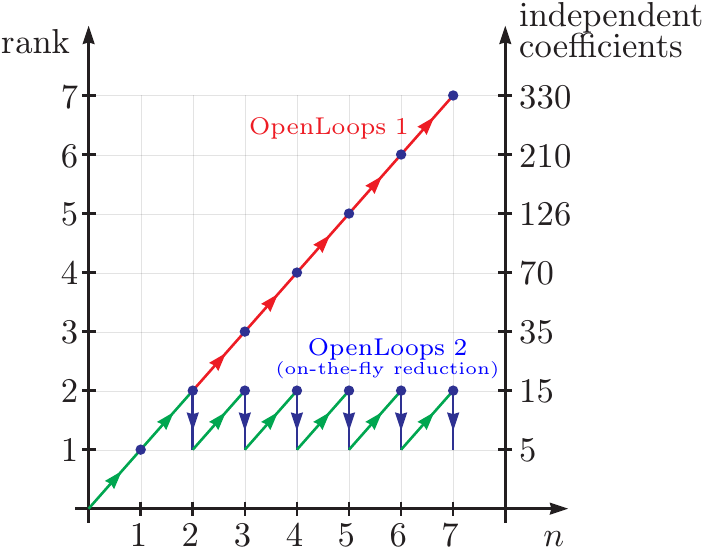}}\; \parbox{2.5cm}{\includegraphics[height=25mm]{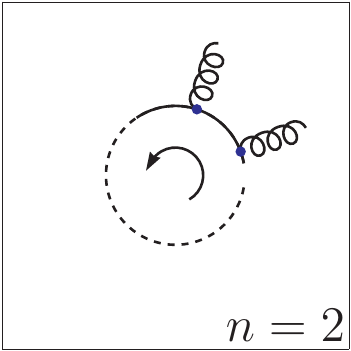}} 
\to\ldots\to \parbox{2.5cm}{\includegraphics[height=25mm]{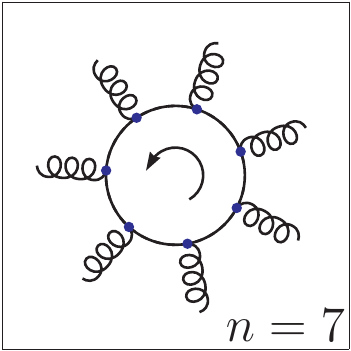}}$
\end{center}\vspace{-2mm}
\caption{Evolution of the tensor rank and number of independent tensor coefficients in \eqref{eq:tcoeff} with the number $n$ of dressed segments for 
the example of a seven-gluon scattering amplitude.
While in \OpenLoops{}~1 \cite{Cascioli:2011va} tensor integrals are only reduced after dressing the diagrams, 
\OpenLoops{}~2 \cite{OL2} allows for an on-the-fly reduction of tensor integrands. 
Here a reduction step is performed after the second dressing step, reducing the rank in $q$ to one, before the third dressing step 
increases the rank again to two. This procedure is continued until the loop is fully dressed.
\label{fig:OL1OL2_r_vs_n} } 
\end{figure}

\noindent
\OpenLoops{} \cite{Cascioli:2011va,OL2} is a fully automated tree-level and one-loop tool for the numerical 
calculation of high-energy scattering amplitudes, which are a key ingredient in multi-purpose Monte Carlo generators. 
The helicity- and colour-summed $n$-particle scattering probability densities
\bea
\calW_{\ssst{LO}}=
\sum\limits_{\hel,\col}
|\calM_{0}|^2,\quad
\calW_{\ssst{NLO}}^{\ssst{virtual}}=
\sum\limits_{\hel,\col} 
2\,\re \Big[\calM_{0}^*\calM_{1}\Big] \quad \text{with}\;
\calM_{l}=\sum\limits_d\calM_{l}^{(d)}
\label{M2W}
\eea
are given by the sums of Feynman diagrams $d$ with $l=0,1$ loops and $n$ external particles.
A one-loop diagram amplitude can be written as
\bea \mathcal{M}^{(d)}_{1}&=& \mathcal{C}^{(d)}_1\,
\int\!\rd^D\momq\, \f{\Tr \Big[{\calN}(q)\Big]}{\Dbar{0}(\momq)\cdots \Dbar{N-1}(\momq)}
\,= \,
\vcenter{\hbox{\scalebox{1.}{\includegraphics[height=20mm]{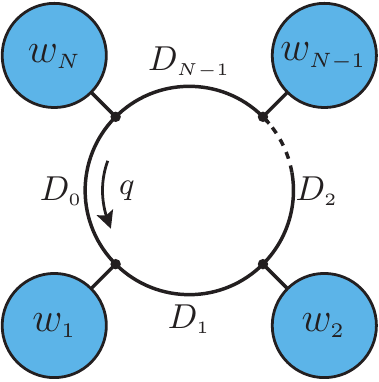}}}}
{}\label{eq:A1d} \eea
with $\Dbar{i}(\momq)=(\momq + \momp{i})^2-\mass{i}^2$ and the colour factor $\mathcal{C}^{(d)}_1$. 
The loop momentum $\momq$ is a vector in $D=4-2\eps$ dimensions (marked by the bar), whereas
the external momenta $\momp{i}$ are four-dimensional. The numerator is constructed by 
cutting the loop open at one propagator $\Dbar{0}$ and dressing the initial open loop $\calN_{0}=\idop$ in recursion steps
\be
\calN_n(q)=\calN_{n-1}(q)\seg_k(q), \qquad n \leq N{},
\label{eq:OLrec}
\ee
exploiting the factorisation of the numerator into segments,
\be
\Tr\Big[\calN(q)\Big]= \Tr\Big[\calN_N(q)\Big]= 
\Big[\seg_1(q)\Big]_{\beta_0}^{\beta_1}\,
\Big[\seg_2(q)\Big]_{\beta_1}^{\beta_2}\cdots
\Big[\seg_{N}(q)\Big]_{\beta_{N-1}}^{\beta_N} \;\delta_{\beta_{N}}^{\beta_{0}}\;
=\;\Tr\left[\vcenter{\hbox{\includegraphics[height=20mm]{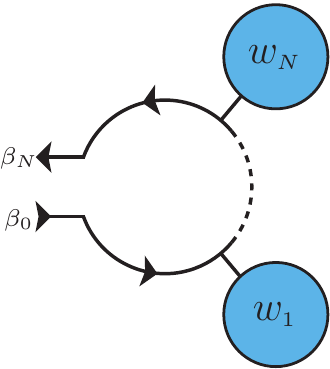}}}\right].
\label{eq:fac}
\ee
The Lorentz or spinor indices $\beta_{0,N}$ of the cut propagator are contracted in the last step. 
A segment consists of a loop vertex and propagator and one or two external subtrees $w_i$,
\be
\Big[\seg_{i}(q)\Big]_{\beta_{i-1}}^{\beta_{i}}
=\quad
\raisebox{3mm}{\parbox{12mm}{
\includegraphics[height=14mm]{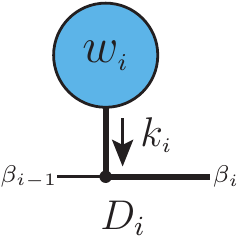}}} \qquad \text{or} \qquad
\Big[\seg_{i}(q)\Big]_{\beta_{i-1}}^{\beta_{i}}
=\quad
\raisebox{3mm}{\parbox{12mm}{
\includegraphics[height=14mm]{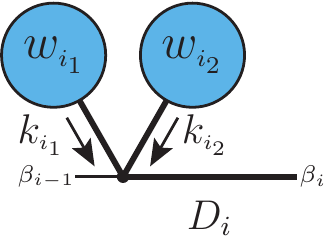}}}{}\quad\quad\;{}.
\label{eq:seg3point}
\ee
Dressing steps are performed numerically for the tensor coefficients $\calN^{(n)}_{\mu_1\dots\mu_r}$ of the partially constructed open loop,
\bea \calN_{n}(q) &=&\raisebox{3mm}{\parbox{65mm}{
\includegraphics[height=14mm]{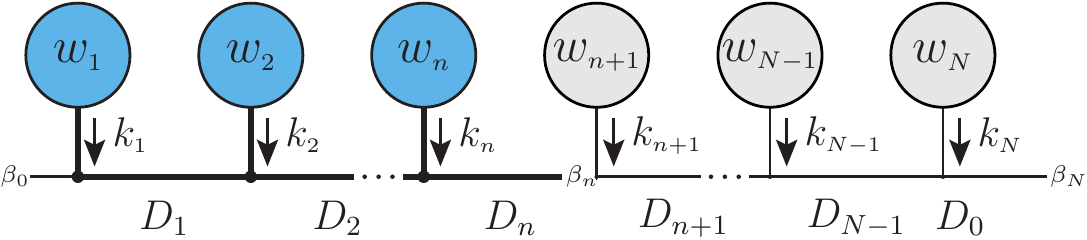}}} =\prod\limits_{i=1}^{k}\seg_i(q)
= \sum\limits_{r=0}^R
\calN^{(n)}_{\mu_1\dots\mu_r}
\,q^{\mu_1}\cdots q^{\mu_r},\qquad \label{eq:tcoeff}\\[-2mm]
& &\underbrace{\phantom{xxxxxxxxxxxxxxx}}_{\text{dressed segments}}\;
 \underbrace{\phantom{xxxxxxxxxxxxx}}_{\text{undressed segments}}\nonumber
 \eea
while the analytical structure in the loop momentum and the scalar denominators  $\Dbar{i}$ is fully retained. Each dressing step increases the 
rank $R$ by zero or one in renormalisable models, using the Feynman gauge.
While in the previous version \OpenLoops{}~1 \cite{Cascioli:2011va} the tensor reduction was performed a posteriori with external libraries, 
such as  \Collier{} \cite{Denner:2016kdg} or \Cuttools{} \cite{Ossola:2007ax},
leading to a high tensor rank and number of independent coefficients, the recently introduced on-the-fly reduction \cite{Buccioni:2017yxi}, 
implemented in \OpenLoops{}~2 \cite{OL2},
keeps the rank and hence the complexity of intermediate results low (see Fig.~\ref{fig:OL1OL2_r_vs_n}).
In the on-the-fly approach operations such as tensor reduction, diagram merging and partial helicity summations are performed interleaved with the 
dressing steps (see \cite{Buccioni:2017yxi}).
\section{The on-the-fly reduction}\noindent
The on-the-fly reduction formulas are based on \cite{delAguila:2004nf} and have the form
\bea
q^\mu q^\nu  &=& \sum\limits_{i=-1}^{3}\lb A^{\mu\nu}_{i} + B^{\mu\nu}_{i,\lambda}\,q^{\lambda} \rb D_i(q) {} \label{eq:qqred}  \quad\text{with}
 \quad D_{-1}(q)=1,
\eea
where the coefficients $A^{\mu\nu}_{i}$ and $B^{\mu\nu}_{i,\lambda}$ are $q$-independent. This is also valid for triangles in renormalisable 
theories \cite{delAguila:2004nf,Buccioni:2017yxi} if we set terms
involving $\Dbar{3}$ and $\momp{3}$ to zero. The loop momentum dependence resides in the reconstructed denominators $D_i=\Dbar{i}-\tilq^2$ which
cancel denominators in the full integrand, leading to four new topologies with pinched propagators. The terms $\propto \tilq^2$, 
where $\tilq=\momq-q$ is $(D-4)$-dimensional, lead to rational terms of type $R_1$ \cite{delAguila:2004nf}.
A reduction step \eqref{eq:qqred} can be applied to the partial integrand of an open loop, before it reaches rank three, due to the factorised 
structure of Feynman diagrams,
\be \left[\f{\calN^{\mu\nu}q_\mu q_\nu}{\Dbar{0}\Dbar{1}\Dbar{2}\Dbar{3}} \right] \prod\limits_{i=k+1}^{N}\f{\seg_i(q)}{\Dbar{i-1}}
=\left[\f{\calN^{\mu}_{-1}q_\mu +\calN_{-1}+\tilde{\calN}_{-1}\tilq^2}{\Dbar{0}\Dbar{1}\Dbar{2}\Dbar{3}} 
+\sum\limits_{i=0}^{3}\f{\calN^{\mu}_{i}q_\mu +\calN_{i}}{\Dbar{0}\cdots\slashed{\Dbar{i}}\cdots\Dbar{3}}\right]
\prod\limits_{i=k+1}^{N}\f{\seg_i(q)}{\Dbar{i-1}}.
\ee
While the complexity due to the tensor rank is kept low throughout the calculation, the creation of pinched topologies potentially 
leads to a huge proliferation of open loops to be processed.
This problem is solved very efficiently by the on-the-fly merging of pinched open loops with open loops of the same 
topology and the same undressed segments, which can also stem from pinches or correspond to lower-point Feynman diagrams 
(for details see \cite{Buccioni:2017yxi,Buccioni:2018zuy}). The final rank-zero and rank-one tensor integrals 
are reduced to scalar box, triangle, bubble and tadpole integrals using integral level identities \cite{delAguila:2004nf,Buccioni:2017yxi},
which are then evaluated with \Collier{} \cite{Denner:2016kdg} or \OneLoop{} \cite{vanHameren:2010cp}.

\section{Numerical stability and any-order expansions} \label{sec:aoexp}
\begin{figure}[t]\begin{center}
  \parbox{0.4\textwidth}{$\includegraphics[width=0.25\textwidth]{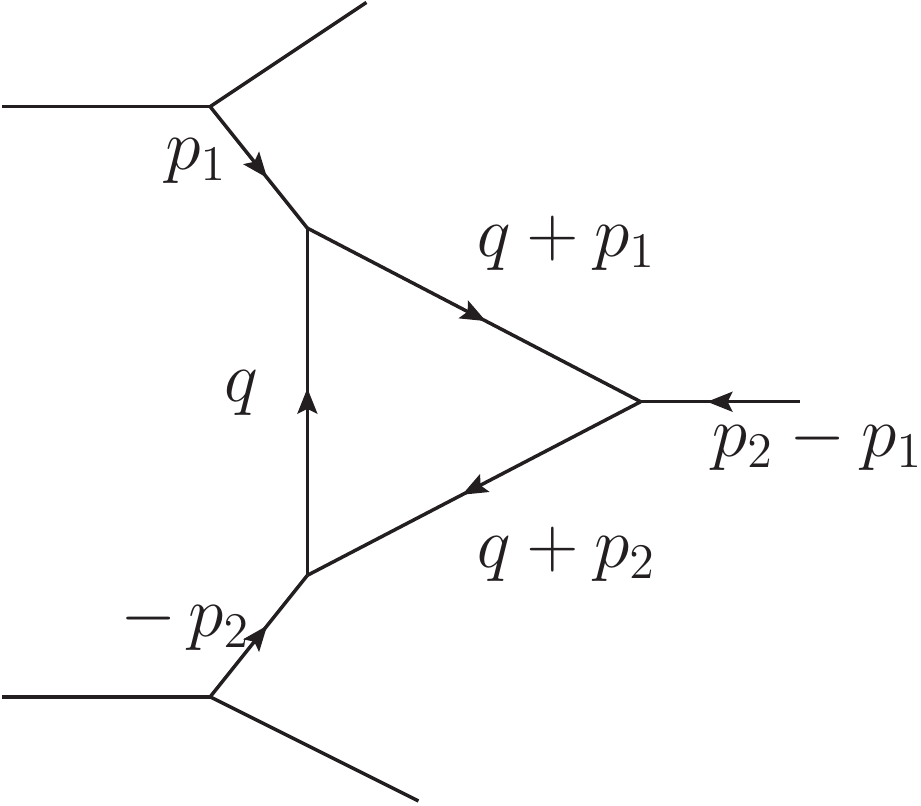}$}
\parbox{0.4\textwidth}{
 \beas
\momp{1}^2 &=& -p^2<0,\\ 
\momp{2}^2 &=& -p^2 (1+\delta),\qquad 0 \leq \delta\ll 1,\\
(\momp{2}-\momp{1})^2 &=& 0\\
\Rig \sqrt{\Delta} &=& \frac{p^2}{2} \delta
\eeas}
\end{center}
\caption{\label{fig:triangle} Triangle configuration leading to Gram determinant instabilities.}
\end{figure}

The main source of numerical instabilities in on-the-fly reduction steps is the appearance of small Gram determinants, especially those 
constructed from two external momenta,
\be
\Delta =- \Delta_{12}=  (p_{1}\cdot p_{2})^2 - p^2_{1} p^2_{2}.
\label{eq:Gram12}
\ee
The three external momenta $p_1,p_2,p_3$ in the $D_{i}$ in \eqref{eq:qqred} do not play equal roles, since only $p_1,p_2$ are used in the 
construction of a basis $l_1,\ldots,l_4$, in which $q^\mu$ is decomposed.
A simple permutation of the propagators in \eqref{eq:qqred},
\bea
\{D_1, D_2, D_3\} \;\longrightarrow\; \{D_{i_1}, D_{i_2}, D_{i_3}\},
\label{eq:propperm}
\eea
such that \bea
\frac{|\Delta_{i_1i_2}|}{Q^4_{i_1i_2}}\;=\; \max\left\{
\frac{|\Delta_{12}|}{Q^4_{12}},\,
\frac{|\Delta_{13}|}{Q^4_{13}},\,
\frac{|\Delta_{23}|}{Q^4_{23}}\right\},\qquad Q^2_{ij}=\max\{|p_i\cdot p_i|, |p_i^2|, |p_j^2|\}
\label{eq:maxgramdet}
\eea
allows us to avoid this type of instability completely in topologies with four or more propagators.
As a result, a single t-channel topology (Fig.~\ref{fig:triangle}) accounts for all rank-two Gram determinant 
instabilities in the hard phase space region.

In this case, we expand the reduction formula \eqref{eq:qqred} (without the $\Dbar{3}$-term) in the small parameter $\delta=\f{2\sqrt{\Delta}}{-p_1^2}$.
Consider for example the rank-one massless-propagator triangle
\bea
C^\mu&=& \int\!\mathrm{d}^D\!q\,\frac{q^\mu}{\Dbar{0} \Dbar{1} \Dbar{2}} =                                                                                        
\frac{1}{\delta }  C_0\left(p_1^2,p_1^2 (1+\delta)\right) \left[-\momp{1}^{\mu}(1+\delta)+\momp{2}^{\mu}\right] \nonumber \\ & &
+\frac{2}{\delta ^2 p^2} \left\{
 B_0\left(p_1^2\right) \left[-\momp{1}^{\mu}(1+\delta)+\momp{2}^{\mu}\right]    
+ B_0\left(p_1^2 (1+\delta)\right) \left[(\momp{1}^{\mu} - \momp{2}^{\mu}) (1+\delta) \right]   \right\}, \label{eq:Cmuexact}
\eea
with 
\be C_0 (\momp{1}^2,\momp{2}^2) = \int\!\mathrm{d}^D\!q\, \frac{1}{\Dbar{0} \Dbar{1} \Dbar{2}}, \qquad
B_0 (\momp{1}^2) =\int\!\mathrm{d}^D\!q\, \frac{1}{\Dbar{0} \Dbar{1}}.\ee
Separating the reduction formula and the master integral evaluation leads to $\f{1}{\delta^n}$-poles in intermediate results
and hence severe numerical instabilities.
If, however, the scalar integrals $B_0$ and $C_0$ are expanded in $\delta$ directly in \eqref{eq:Cmuexact}, these poles cancel completely,
\bea
C^\mu &=& \f{\momp{1}^{\mu}+\momp{2}^{\mu}}{2 p^2} \left[- B_0(p_1^2)  +1 \right]    
 + \delta\; \f{\momp{1}^{\mu}+2\momp{2}^{\mu}}{6 p^2} \left[ B_0(p_1^2)  \right]    +\mathcal{O}(\delta^2).\label{eq:Cmuexptruncated}
\eea
The same behaviour was found for tensor rank up to three in all cases with massless and massive propagators relevant for QCD \cite{Buccioni:2017yxi}. 

A disadvantage of this procedure is that the truncation of the expansion at a fixed order spoils the rescaling test used to 
estimate the numerical uncertainty of the calculation \cite{Cascioli:2011va,Buccioni:2017yxi}.
In order to avoid this issue we developed and implemented an any-order expansion. Since the $\f{1}{\delta^n}$-poles in \eqref{eq:Cmuexact} cancel, 
we can substitute
\be
\f{1}{{\delta}^n} B_0(p_1^2(1+{\delta})) \to B_{0,n}(p_1^2,{\delta}), 
\qquad \f{1}{{\delta}^n} C_0(p_1^2,p_1^2(1+{\delta})) \to C_{0,n}(p_1^2,{\delta})\\[-3mm]
\ee
with
\bea
 B_{0,n}(p_1^2,{\delta}) &=& \,
\sum\limits_{m=n}^{\infty} {\delta}^{m-n}\left[\f{1}{m!}\lb \f{\p}{\p\delta} \rb^m  
B_0\left(p_1^2 (1+{\delta})\right)\right]_{ \delta=0}\\
 C_{0,n}(p_1^2,{\delta}) &=& \,
\sum\limits_{m=n}^{\infty} {\delta}^{m-n}\left[\f{1}{m!}\lb \f{\p}{\p\delta} \rb^m  
C_0\left(p_1^2,p_1^2 (1+{\delta})\right)\right]_{ \delta=0}.
\eea
Our rank-one example \eqref{eq:Cmuexact} then takes the form
\bea
C^\mu
&=& (\momp{1}-\momp{2})^{\mu} \left[ \f{B_{0,1} +2\,B_{0,2}}{p^2}
- C_{0,1} \right]+ \momp{1}^{\mu} \left[ \f{B_{0,1}}{p^2} - C_{0} \right].
\eea
Closed formulas were derived and implemented for $\lb \f{\p}{\p\delta}\rb^m\!\! B_0$ and $\lb \f{\p}{\p\delta}\rb^m\!\! C_0$ for all
topologies relevant in QCD, and are applied order by order until the relative truncation error becomes smaller than the target precision of $10^{-16}$ or $10^{-32}$ 
in double or quadruple precision calculations respectively.
In this way, the truncation error is avoided entirely and the rescaling test of the numerical accuracy is valid. 
The details of the any-order expansion will be discussed in \cite{OL2_stability}.

\begin{figure}\begin{center}
\begin{tabular}{cc}
\includegraphics[width=0.4\textwidth]{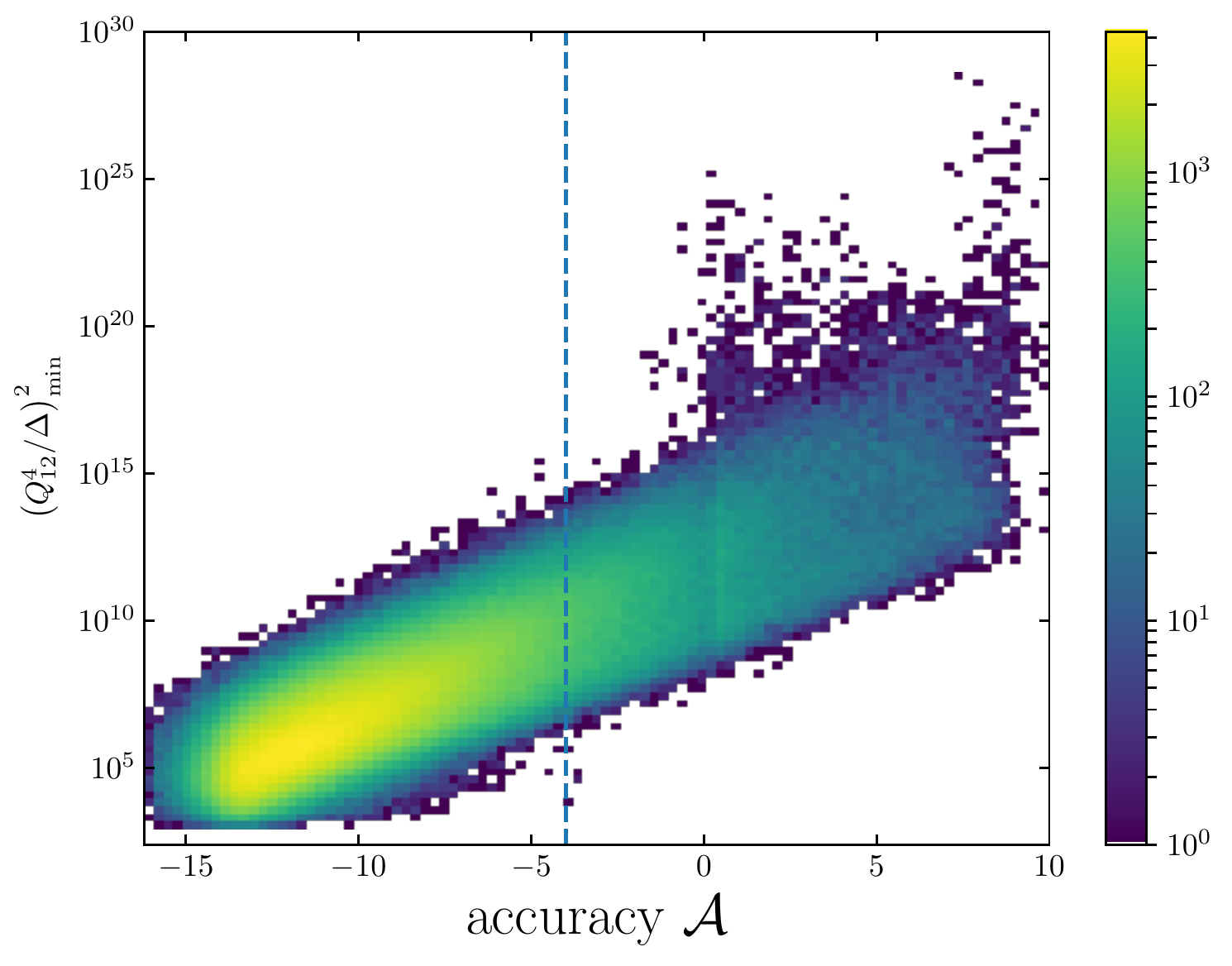} 
&
\includegraphics[width=0.4\textwidth]{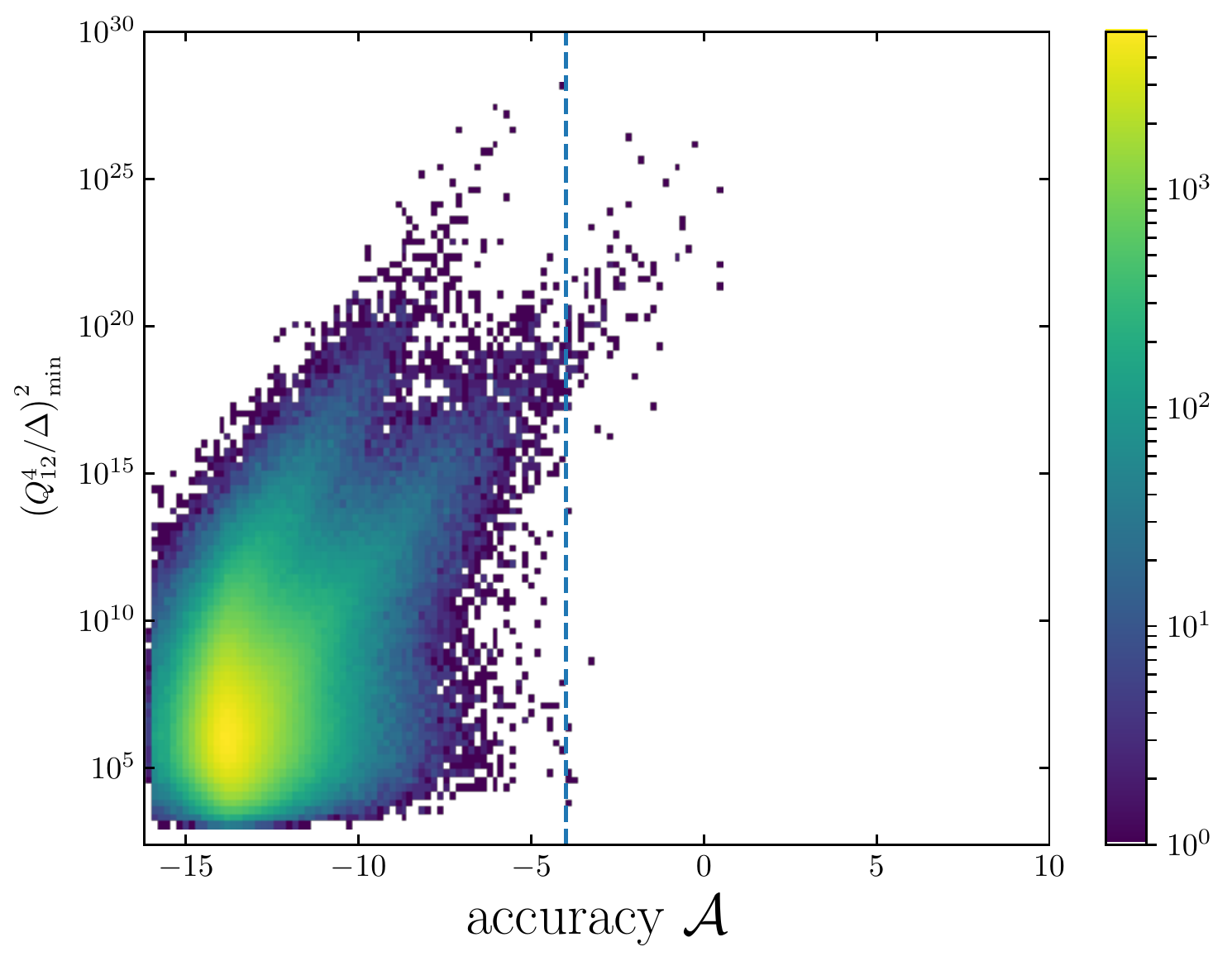} 
\\{(a)}&{(b)} \\
\includegraphics[width=0.4\textwidth]{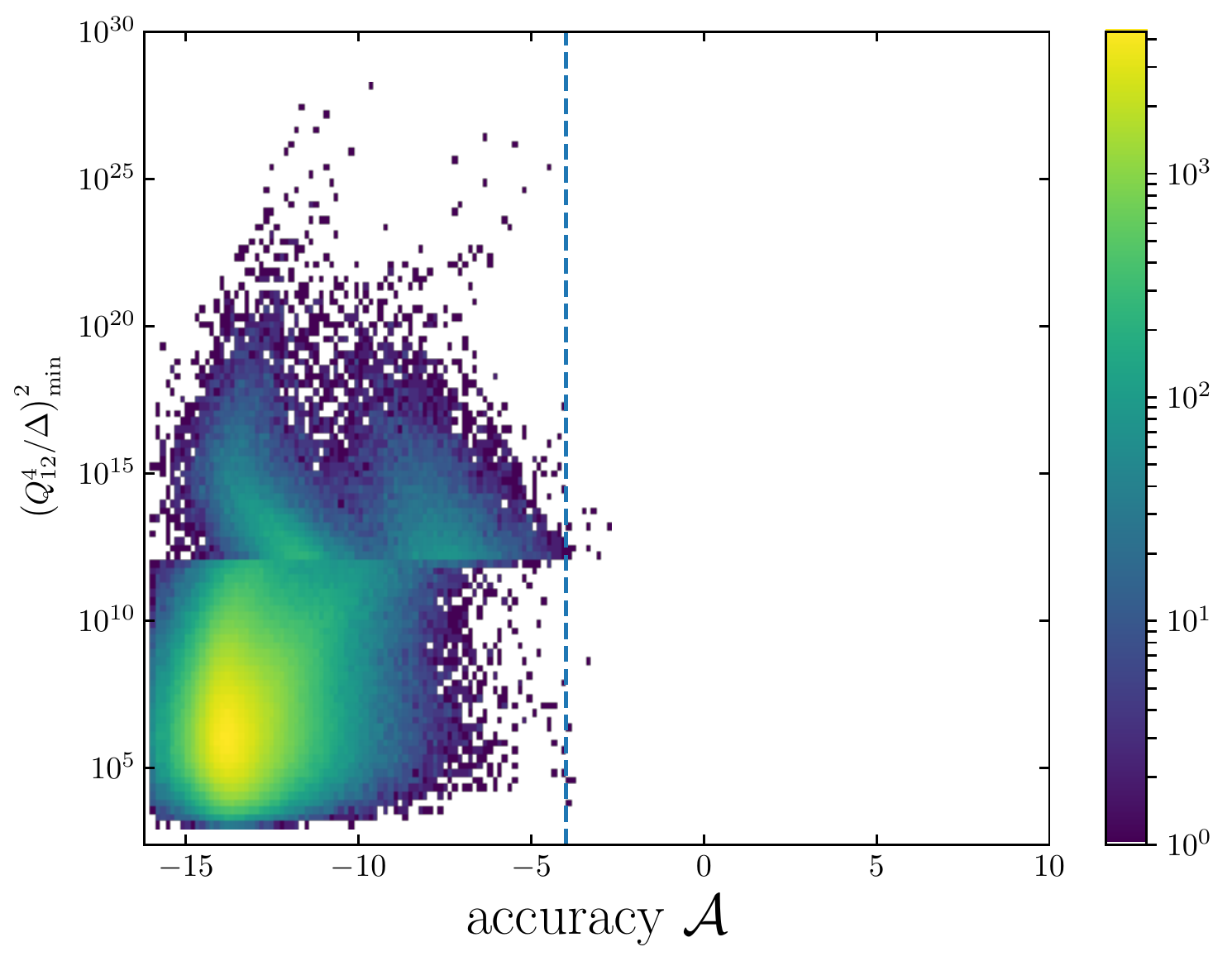}
&
\includegraphics[width=0.4\textwidth]{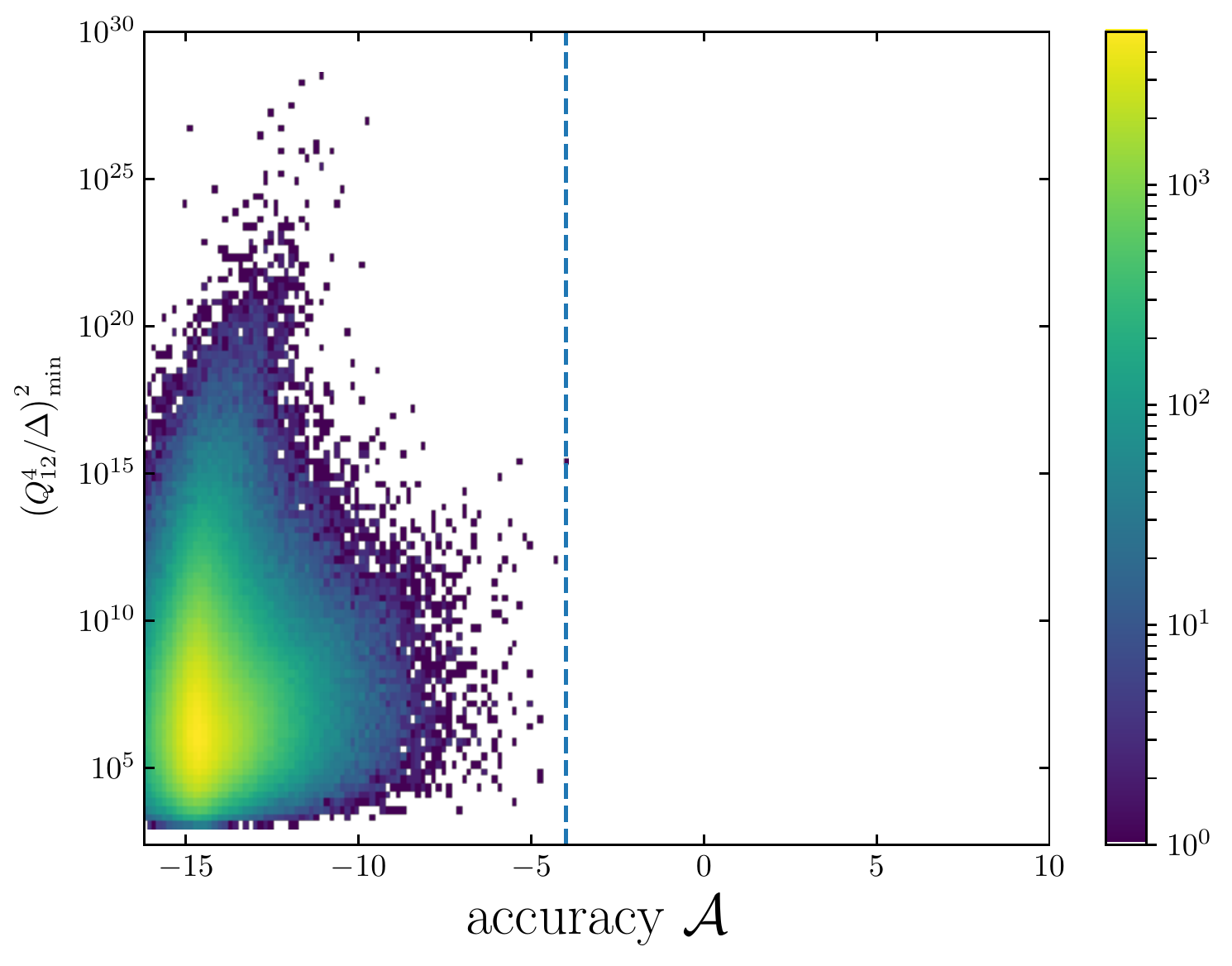}
\\{(c)}&{(d)} 
\end{tabular}\end{center}
\caption{\label{fig:correlationplots} 
Correlation between the instability $\calA$ in double precision and the largest
$(Q_{12}^4/\Delta)^2$ in the event, with any rank-two Gram determinant $\Delta$
and corresponding $Q_{12}^2$ from \eqref{eq:maxgramdet}, in a sample of $10^6$ phase space points for $\mathrm{g g} \to \mathrm{t\bar{t}gg}$.
The on-the-fly reduction was used
without special treatment of Gram determinants in (a), and with the permutation trick \eqref{eq:propperm} 
and analytic expressions for triangle reduction in (b). 
In addition to these improvements, the fixed-order Gram-determinant expansions up to $\delta^2$ for $\delta<\deltathr=10^{-3}$ were used in (c), 
and the any-order expansion in (d).}
\end{figure}
In order to illustrate the effect of the various improvements in the on-the-fly reduction of \OpenLoops{}~2, we show the 
correlation between the instability and the smallest rank-two Gram determinant $\Delta$ in a calculation, taking a sample of $10^6$ random phase space points for 
the process $\mathrm{gg} \to \mathrm{t\bar{t}gg}$.
The results without any special treatment of Gram determinants in Fig.~\ref{fig:correlationplots} (a) show a strong correlation
with rank-two Gram determinants over twenty orders of magnitude. In fact, we observe a  quadratic or faster scaling in
$Q_{12}^4/\Delta$, consistent with the explicit $\Delta$-dependence of the coefficients in \eqref{eq:qqred}. These results
are stabilised using the permutation trick \eqref{eq:propperm} and analytic expressions, such as \eqref{eq:Cmuexact} for the triangle reduction, 
as shown in in Fig.~\ref{fig:correlationplots} (b). 
Adding fixed-order Gram-determinant expansions, such as \eqref{eq:Cmuexptruncated}, up to $\delta^2$,
introduces a horizontal discontinuity in Fig.~\ref{fig:correlationplots} (c) due to the threshold, 
below which the exact formula is replaced by the expansion.
Using the any-order expansion instead leads to an extremely stable result which does not suffer from a threshold discontinuity as shown in 
Fig.~\ref{fig:correlationplots} (d).

\begin{figure}[t]\begin{center}
\includegraphics[width=\textwidth]{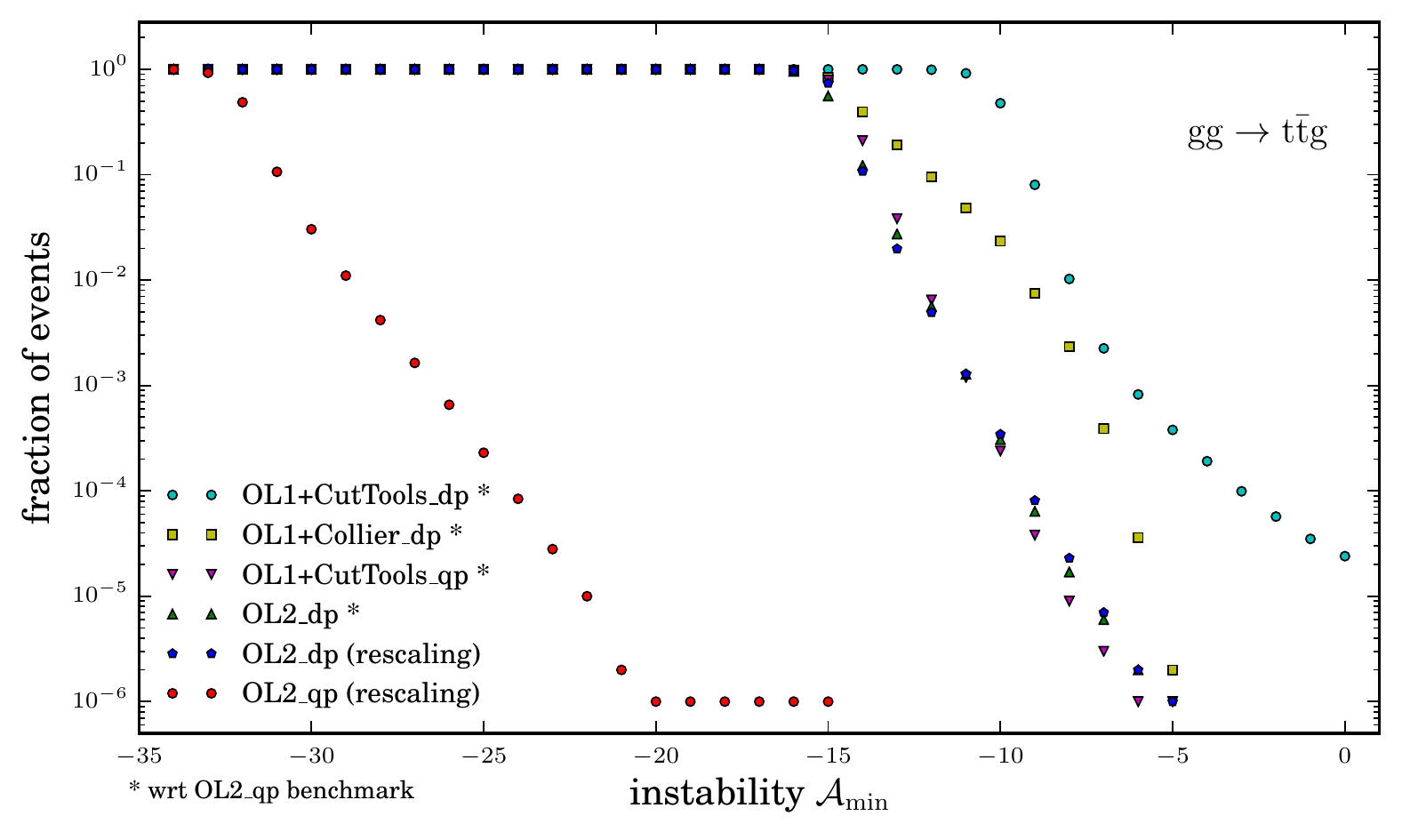}
\end{center}\vspace*{-5mm}
\caption{Stability distributions for a sample
$2 \to 3$  process. \label{fig:numstab_2to3}}
\end{figure}

We now present numerical stability studies for the on-the-fly algorithm in various modes of \OpenLoops{}.
In Fig.~\ref{fig:numstab_2to3} we show the fraction of points with an accuracy $\calA<\calA_{\ssst{min}}$ 
plotted against $\calA_{\ssst{min}}$ for a sample of $10^6$ homogeneously distributed random 
phase space points at $\sqrt{s}=1$\,TeV, for the process $\mathrm{g g} \to \mathrm{t\bar{t}g}$.
Infrared regions are excluded through cuts, $p_{i,\mathrm{T}} >50$\,GeV and $\Delta R_{ij} > 0.5$ for massless final-state partons.

The numerical accuracy of the double precision (dp) results is defined w.r.t.~a quadruple precision (qp) benchmark, 
$\calA = \log_{10} \left|(\calW_{\ssst{dp}}-\calW_{\ssst{qp}})/\min\left\{|\calW_{\ssst{dp}}|,|\calW_{\ssst{qp}}|\right\}\right|$.
The qp benchmark is derived in \OpenLoops{}~2 with the on-the-fly reduction (OL2), which is fully implemented in dp and qp, including 
all stability improvements.
The final scalar integrals are evaluated with \OneLoop~3.6.1 \cite{vanHameren:2010cp} in qp. The accuracy of the benchmark OL2\_qp 
is determined with the rescaling test.
This is the first full qp implementation of a one-loop amplitude generator, yielding up to $32$ correct digits. 
For this process only one out of a million phase space points 
has less than $20$ correct digits. This is in contrast to \OpenLoops{}~1 (OL1) with \Cuttools{}~1.9.5 \cite{Ossola:2007ax} 
in qp, where dp-contamination inside \Cuttools{} prevents more than $16$ correct digits.
The scalar integrals for OL2 calculations in dp were evaluated with \Collier{}~1.2 \cite{Denner:2016kdg}.

For this process, OL2 in dp yields an improvement of $1-3$ orders of magnitude in numerical accuracy as compared 
to OL1+\Collier{} in dp, which in turn gains many orders of magnitude compared to OL1+\Cuttools{} in dp, especially in the tail,
where the latter becomes unreliable. In fact, the accuracy of OL2 in dp is comparable to OL1+\Cuttools{} in qp, 
which used to be the benchmark in \OpenLoops{}~1. 
The accuracy of OL2 was measured once against the OL2\_qp benchmark and once with the rescaling test, 
the results of which are in excellent agreement.
In \OpenLoops{}~2 we use a stability rescue system for dp calculations similar to the one in \OpenLoops{}~1, 
which is based on the rescaling test in dp and a re-run in qp for phase space points, for which a target 
accuracy is not reached. 

\section{CPU efficiency} \label{sec:speed}
\begin{figure}[t]\begin{center}
\includegraphics[width=\textwidth]{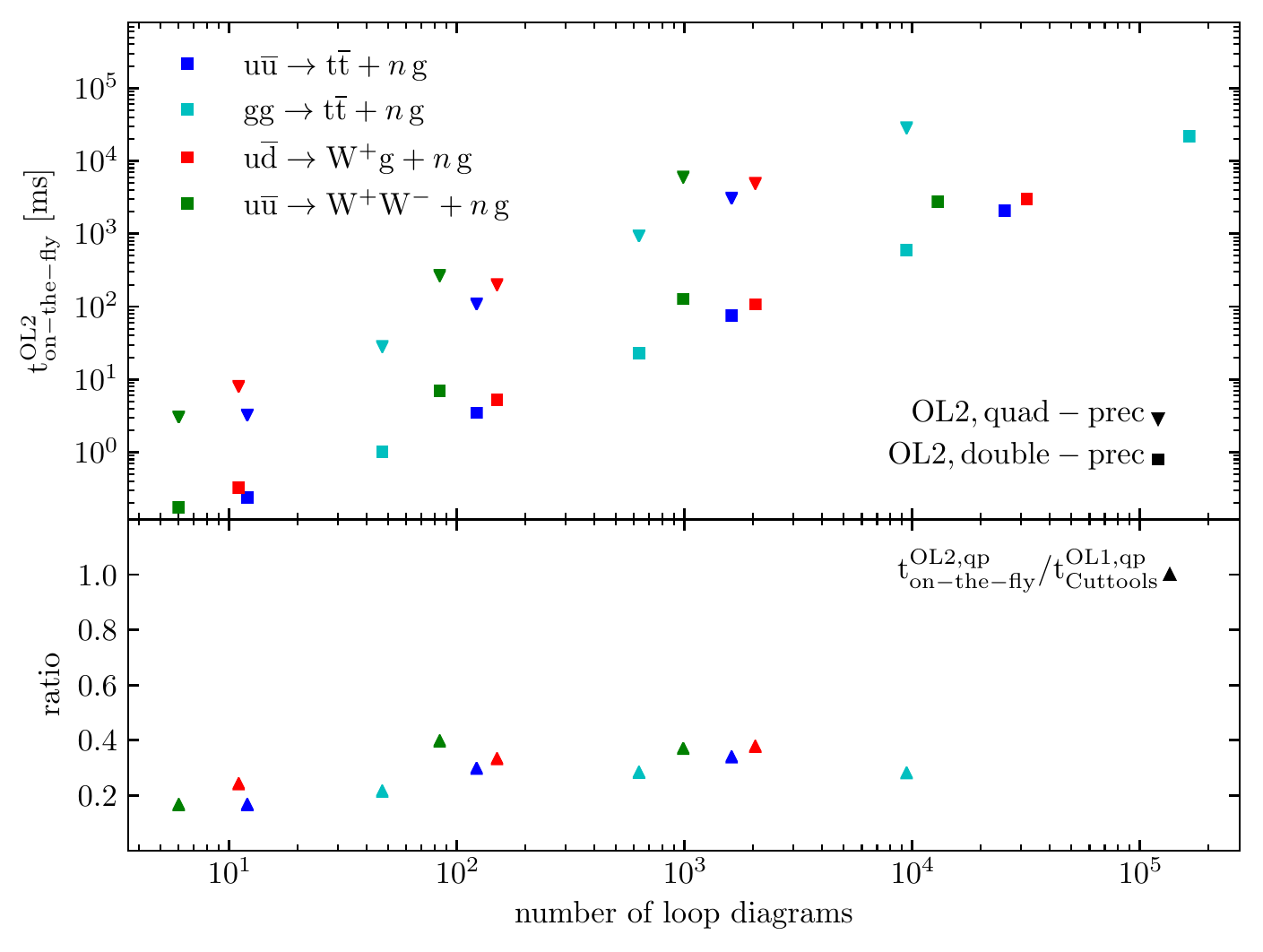}
\end{center}\vspace*{-5mm}
\caption{CPU runtimes per phase space point on a single $\text{Intel}$ i7-4790K core with gfortran-4.8.5
for a one-loop scattering probability density plotted versus the number of one-loop diagrams
 in double (dp) and quadruple precision (qp). 
From left to right, the number of additional gluons is $n=0,\dots,3$ in dp and $n=0,\dots,2$ in qp.
In the lower frame the ratio of the runtime in \OpenLoops{}~2 (OL2) in qp to the runtime in 
\OpenLoops{}~1+\Cuttools{} (OL1) in qp is presented.\label{fig:speed_qp}}
\end{figure}
In this section we briefly discuss the speed of the on-the-fly algorithm. The upper frame in Fig.~\ref{fig:speed_qp} 
shows the runtime versus the number of one-loop Feynman diagrams
for four $2\to 2+n$ process classes with $n=0,\dots,3$ additional gluons in the final state. We find in good approximation, 
that the order of magnitude of the runtime scales linearly with the order of magnitude 
of the number of one-loop diagrams. Computing a phase space point in qp takes a factor $20-80$ longer than in dp, 
depending on the process. 
In high-multiplicity processes, this factor tends to be larger than in simple $2\to 2$ processes.
The lower frame shows the ratio between the OL2 runtime in qp to the OL1+\Cuttools{} runtime in qp, 
where we find a gain in speed of a factor $3-5$ due to the on-the-fly method. Hence, the efficiency improvement in \OpenLoops{}
due to the on-the-fly method is even more pronounced in qp
than in dp, where we found a speed-up of a factor $2-3$ \cite{Buccioni:2017yxi}. 
      
\section{Conclusion}

We have presented the on-the-fly method for the automated calculation of scattering
amplitudes at one loop. Exploiting the factorised structure of open loops in a systematic way, we perform 
operations, such as tensor reduction, helicity summation and diagram merging, on-the-fly during the open-loop recursion.
This approach reduces the complexity of intermediate results and operations significantly, 
leading to a substantial gain in CPU efficiency.

The on-the-fly integrand reduction allows us to isolate Gram determinant instabilities 
in triangle topologies with a particular kinematic configuration and to cure them by means of simple analytic expansions,
which can be performed to any order. With a small set of simple optimisations the on-the-fly algorithm achieves 
an unprecedented level of numerical stability, which is a particularly attractive feature for the calculation of 
real--virtual contributions at NNLO.

This algorithm is fully implemented in double and quadruple precision, and validated for a wide range of SM 
processes at NLO QCD. It will become publicly available in the upcoming release of \OpenLoops{}~2.

\subsection*{Acknowledgements}
This research was supported in part by the Swiss National Science Foundation
(SNF) under contracts PP00P2-128552 and BSCGI0-157722.

\providecommand{\href}[2]{#2}\begingroup\raggedright\endgroup

\end{document}